\documentstyle[twocolumn,aps]{revtex}
\draft

\begin{document}

\title{Observation of Scarred Modes in Asymmetrically Deformed Microcylinder Lasers}

\author{Sang-Bum Lee, Jai-Hyung Lee, and Joon-Sung Chang}
\address{Condensed Matter Research Institute, School of Physics, Seoul National University, Seoul 151-742, Korea}
\author{Hee-Jong Moon}  
\address{Department of Optical Engineering, Sejong University, Seoul 143-747, Korea}
\author{Sang Wook Kim and Kyungwon An}
\address{Center for Macroscopic Quantum-Field Lasers, Department of Physics, Korea Advanced Institute of Science and Technology, Taejon 305-701,Korea}

\date{\today}

\maketitle

\begin{abstract}
We report observation of lasing in the scarred modes in an asymmetrically deformed microcavity made of liquid jet. The observed scarred modes correspond to morphology-dependent resonance of radial mode order 3 with their $Q$ values in the range of $10^6$.  Emission directionality is also observed, corresponding to a hexagonal unstable periodic orbit.

\end{abstract}

\pacs{PACS number(s): 05.45.Mt, 42.55.Sa}

The morphology-dependent resonance (MDR) modes, also known as whispering gallery modes, in a sphere or a cylinder can support ultrahigh cavity quality factor $Q$ in the optical region due to repetitive total reflections of light at the circular boundary. In recent years circular microcavity lasers with microdroplets, microcylinders, and microdisks have been widely investigated for the purpose of developing ultra-efficient optical devices utilizing such ultrahigh $Q$ as well as the gain enhancement originating from the cavity quantum electrodynamics \cite{Campillo91,Lin92}. In addition, deformed microcavities have drawn much interest due to their output emission directionality, which can be practically important in designing efficient microlasers and light-emitting diodes. 
N\"{o}ckel {\em et al.} have studied the effects of deformation with a ray-optics model \cite{Nockel94,Mekis95,Nockel96} along with the wave equation analysis \cite{Nockel97} in an asymmetric cavity of 2D quadrupolar deformation described by the equation $r(\phi) = a(1 + \eta \cos 2 \phi)$ with $a$ the undeformed radius, $\eta$ the deformation parameter and $\phi$ the polar angle.  In their study deformation-induced $Q$ spoiling was predicted along with dynamical localization and chaos-assisted tunneling.
Gmachl {\em et al.} demonstrated emission directionality and enhancement in emission power in a highly deformed semiconductor microcavity  \cite{Gmachl98}.  The emission directionality in this case
came from the so-called bow-tie shape orbits of ray, which correspond to the {\em stable} resonance islands in the Hamiltonian dynamics. 

The bow-tie modes, however, cannot exist in cavities of low index of refraction ($< 2$) such as optical fiber or cylindrical dye jets. For these cavities of low index of refraction, neither {\em stable} resonance islands nor {\em stable} periodic orbits exit in ray optics model. Any directional emission would come from the {\em chaotic} MDR modes \cite{Nockel96,Nockel97} without real mode structures in the lasing spectrum. Even in these cavities, however, a dense set of unstable periodic orbits (UPO's) are still embedded in the chaotic orbits. Although UPO's are found with zero probability in the classical dynamics, in quantum mechanics they manifest themselves in the eigenstates of the system.  There exist extra and unexpected concentrations, so-called {\em scars}, of eigenstate density near UPO's.  

The existence of the scar in general was discussed in both linear \cite{Heller84,Bogomolny88,Berry89} and nonlinear semiclassical theories \cite{Kaplan98}.  Experimentally it has been shown in 
a quantum well in a high magnetic field \cite{Wilkinson96} and Sinai-billiard-shaped microwave cavities \cite{Sridhar91}. So far, no experimental evidence exists for the existence of scars in optical microcavities. If scarred modes can exist in a deformed microcavity, they could also give rise to emission directionality with mode-like structures in the emission spectrum.  Furthermore, the $Q$ values of these modes are expected to be still very high, although degraded from their original ultrahigh values, since the scarred mode corresponds to a ray which undergoes total reflections in its dynamics. The bow-tie mode, on the other hand, corresponds to a ray simultaneously reflecting and refracting off the cavity boundary, and thus its $Q$ tends to degrade severely. 

In this Letter we report the first observation of the lasing in the scarred modes in an asymmetrically deformed microcylinder. The observed scarred modes correspond to MDR of mode order 3 with their $Q$ values as high as $2\times 10^6$, which is 100 times larger than the $Q$ value expected from the refractive escape of the chaotic MDR modes \cite{Nockel96,Nockel97}. From the directionality of the emission, we found that the scarred modes correspond to a hexagonal UPO.

Our experimental setup, similar to that of Ref. \cite{Moon97}, is shown in Figure \ref{setup}. The basic idea of the experiment is as follows. A liquid jet containing fluorescent dye molecules is laterally forced so that a 2D microcavity with {\em variable} quadrupole deformation is created. By adjusting the lateral force we can control the degree of deformation accurately. The cavity is optically pumped that any persisting MDR modes with low loss would undergo laser oscillations. From the lasing spectrum, we can get information on the $Q$ values and the mode orders of the lasing MDR's. 

In details, we used a liquid jet ($a\sim 14 \mu$m), directed upward, from a glass orifice. The liquid in the jet was ethanol doped with Rhodamine B (RhB) at the concentration of $1\times10^{-4}$ M/L (refractive index $n$=1.361). The jet stream was illuminated by a $Q$-switched Nd:YAG pump laser (532 nm, 10 ns) perpendicularly at the position of 3 mm from the orifice.  The polarization of pump laser was parallel to the stream direction, corresponding to TM polarization. Argon gas through a glass tube of 1.0 mm diameter was blown normal to the jet and at the same time perpendicularly to the incident pump laser. The distance between the tip of the tube and the jet was 2 mm. The diffraction patterns of the pump beam created by the jet were projected on a screen and the images were recorded on a CCD camera. From the diffraction patterns we could find the deformation parameter $\eta$, which is defined as $\eta \equiv 1-b/a$, where $2b$ is the length of the minor axis of the deformed jet.
The spectrum of MDR lasing light emitted in the same direction as the gas flow was measured with a spectrometer equipped with a photo-diode array (PDA) on the image plane of the spectrometer. 

Figure \ref{spectrum} shows the typical MDR lasing spectra emitted from the deformed jet for  various $\eta$ values. When the argon gas was not blown to the jet ($\eta$= 0), two groups of repeated MDR peaks appeared with one centered around 608 nm and the other spread over the 620-645 nm range as shown in Fig. \ref{spectrum}(a). The MDR modes in a cylinder are characterized by the mode number or the angular momentum index $l$, the mode order or the radial index $\nu$, and the polarization (TE or TM). From the polarization analysis we found that all of these peaks were TM modes. Based on the fitting described below we also found that the mode orders are $\nu=5$ for the MDR group around 608 nm and $\nu=4$ for the other group around 620-645 nm, respectively. 

The absorption quality factor of the gain medium, $Q_{\rm abs}$, is defined as $Q_{\rm abs}\equiv 2\pi n / \lambda \alpha$, where $\alpha$ is the absorption coefficient and $n$ the refractive index of the dye solution. We found that $Q_{\rm abs} =5\times10^6$, $5\times10^7$, at 608 nm and 630 nm, respectively, from the absorption measurement.
Output coupling efficiency or mode visibility is characterized by the ratio $Q/Q_{\rm leak}$, where $Q \equiv (Q_{\rm leak}^{-1} + Q_{\rm abs}^{-1})^{-1}$ is the total quality factor, and $Q_{\rm leak}$, corresponding to overall cavity leakage loss, is defined as $Q_{\rm leak}\equiv (Q_0^{-1} +Q_{\rm surf}^{-1}+ Q_{\rm def}^{-1} )^{-1}$.
Here $Q_0$ represents the cavity leakage of an ideal undeformed cavity, calculated from the Lorentz-Mie theory using only the real part of $m$ \cite{Barber90}.
It depends weakly on the mode number $l$ and strongly on the mode order $\nu$. 
The term $Q_{\rm surf}^{-1}$ represents the loss induced by microscopic surface imperfection such as surface scattering. The last term $Q_{\rm def}^{-1}$ represents the leakage induced by the cavity deformation alone. For $\eta$=0, this term vanishes. 
Since MDR's of $\nu=4, 5, 6$ have $Q_0\sim 10^9 , 10^7, 10^5$, respectively, and since the MDR group centered around 608 nm can be fitted with $Q_{\rm leak}$ of $2\times10^6$ as discussed below, this group must be of $\nu$=5 mode order with $Q$ degradation possibly by surface imperfection.
Since $Q/Q_{\rm leak}$ for $\nu$=5 modes is of order of unity, the corresponding peaks appeared strongly in Fig.\ \ref{spectrum}(a), while $\nu$=4 modes appeared weak due to their small mode visibility ($Q_{\rm leak}> 10^8$).

As the deformation parameter $\eta$ increased, individual MDR peaks shifted to red. This red shift is mainly due to the increase in the cavity perimeter in the deformation preserving the cross sectional area of the jet, requiring  the wavelength of MDR to increase accordingly \cite{Nockel97}.  The cross sectional area of the jet  increased slightly in the actual experiment due to the drag force exerted by the argon gas flow.  
We could, however, subtract out the contribution from the area increase by measuring the amount of the red shift at the position where the jet returned to a round shape, well above ($\sim$ 3 mm) the deformed region on the jet, with and without the argon flow.   The difference between these two measurements is the contribution from the area increase. Figure \ref{eta-dependence}(a) shows the amount of red shift measured in this way versus the deformation parameter $\eta$, which were independently measured from the diffraction data. The experimental results are in good agreement with the theoretical curve, with no fitting parameter, assuming the quadrupole deformation preserving the cross sectional area.  Based on this agreement we can assure that our estimates on $\eta$ in Fig.\ \ref{spectrum} are fairly good.

As $\eta$ increased, both the $\nu$=5 and $\nu$=4 mode {\em groups} shifted to blue. Such blue shifts of MDR groups are associated with $Q$ spoiling, as observed in the dye-doped microdroplets by Lin {\em et al.} \cite{Lin92}. When the cavity $Q$ is high enough, the main loss in the lasing comes from the gain-medium absorption, and thus the lasing occurs at the wavelength where the absorption is weak, i.e., at the red. One the other hand, when $Q$ is degraded severely, the cavity loss accounts for the major loss in the lasing, and thus the lasing occurs where the emission is strong, i.e., at the blue.  Therefore, by monitoring the blue shifts of MDR groups, one can quantify the $Q$ spoiling caused by the cavity deformation. 
In Fig.\ \ref{spectrum}, as $\eta$ increased beyond 4.7\%, $\nu$=5 modes disappeared completely while the $\nu$=4 mode group showed continuous blue shift until $\eta$ reached 8.8\%. The $\nu$=4 mode group disappeared completely when $\eta$=10.5\%, and instead the mode group corresponding to $\nu = 3$ newly appeared and showed continuous blue shift as $\eta$ increased further. 

In order to quantify such mode-order-dependent $Q$-spoiling in an asymmetrically deformed microcavity, 
we consider a dye laser model adapted for our experimental configuration \cite{Moon-PRL00}.
Let  $N_0$ and $N_1$ be the number densities of dye molecules in the ground and in the first excited electronic singlet state, respectively. The total number of molecules per unit volume is $N_t=N_0 +N_1$. The threshold condition can be written as 
\begin{equation}
N_1 \sigma_e (\lambda) \ge 2\pi n / \lambda Q_{\rm leak} + N_0 \sigma_a (\lambda) \;,
\end{equation}
where  $\sigma_e (\lambda)$ and  $\sigma_a (\lambda)$ are the emission and the absorption cross section of the dye molecule, respectively.  
Eq.\ (1) leads to the minimum fraction $\gamma(\lambda)$ of molecules for laser oscillation as 
\begin{equation}
\gamma (\lambda)  = \frac{N_1}{N_t}=\frac{ 2\pi n /(\lambda N_t Q_{\rm leak}) + \sigma _a  (\lambda)}
{\sigma_e (\lambda) +\sigma_a (\lambda)} \;.
\end{equation}

Fog a given $Q_{leak}$, $\gamma(\lambda)$ curve has a minimum. From the center wavelength of a group of MDR peaks of the same mode order, we can find $Q_{leak}$ associated with this MDR group.
Figure \ref{eta-dependence}(b) summarizes the results of fitting, showing the fitting $Q_{\rm leak}$ value of each mode-order group at various $\eta$ values. 
Our results clearly demonstrate that the $Q$ spoiling strongly depends on the mode orders: The MDR's of the lower mode order are the more robust to cavity deformation.  Since the MDR of low mode order have relatively large initial incident angle in the ray optics model, more deformation is required in order to make the ray associated with the mode escape refractively.
It should be pointed out, however, that no resonance modes can exist for $\eta >$ 12\% according to the ray optics model (see Fig.\ \ref{poincare sos}(b)). Our result for $\eta\sim$ 14.3\% as shown Fig.\ \ref{spectrum}(i) can be explained only by the wave nature of MDR, i.e., the scars. 

To illustrate the ray dynamics in a deformed cavity, we present the ray motion in phase space using Poincare surface of section with Birkoff coordinates, in which every time a ray collides with the cavity boundary both the azimuthal angle $\phi$, at which it hits, and its angle of incidence $\chi$ with respect to the boundary are recorded. Figure \ref{poincare sos}(b) shows the Poincare surface of section (PSOS) with $\eta$ = 14\%, where it is shown that no stable island exists except the bifurcated period-4 island on which the ray escapes refractively after a few collisions.

The wave nature of MDR in a asymmetrically deformed cavity can be seen in the plane wave scattering problem for the same geometry as that of our experimental setup \cite{Barber90,Narimanov99}. 
We calculated the {\em scarred} mode as shown in Fig.\ \ref{scar}(a). The concentrated intensities of the field, resembling that of whispering gallery modes, are clearly seen near the boundary. For comparison the non-resonant case is also given in Fig.\ \ref{scar}(b), where the enhanced intensities near the boundary are no longer recognized except for the common diffracted waves of the incident plane wave. We emphasize that no stable periodic orbit in ray dynamics can exist at this high deformation as shown in Fig.\ \ref{poincare sos}(b). 

The existence of the scarred MDR modes was also investigated in the far-field distribution of our deformed microlaser in the direction to which the emission spectra were taken in Fig.\ \ref{setup}. The result is shown in Fig.\ \ref{direction}. For small $\eta$ the central region of the distribution is enhanced, which can be ascribed to the chaotic refractive emission of high order MDR's from the highest curvature points at $\phi=0^\circ$ and $180^\circ$ \cite{Nockel96,Nockel97}. For intermediate deformations, the distribution has two peaks around $\pm 10^\circ$, indicating the major portion of emission comes from the points where the curvature is not highest ($\phi\sim 10^\circ , 170^\circ$), possibly due to the dynamical eclipsing \cite{Chang00}.  Such emission pattern is impossible for an ellipse, for which the maximal emission has to occur at the highest curvature points, $\phi=0^\circ$ and $180^\circ$.  
For higher deformation, $\eta > $10\%, two new peaks appear around $\pm 24^\circ$ while the peaks around $\pm 10^\circ$ fade out. These peaks exhibit the mode-like spectrum as shown in Fig.\ \ref{spectrum}, corresponding to MDR's of $\nu=3$.
In Fig.\ \ref{poincare sos}(a) we show the PSOS with $\eta =$10\%, in which the dotted line is the adiabatic curve \cite{Nockel97} for $\nu=3$ mode, representing the ray dynamics of $\nu=3$ mode in the ray optics picture, corresponds to a hexagonal orbit described by the period-6 resonance island at $\eta=$ 10\%. 
Since the ray on this resonance island is totally reflected off the boundary, the emission can be caused only by the evanescent leakage at the corners of the hexagon, resulting in the observed emission angles. Although at $\eta=$ 14\% this hexagonal orbit loses its stability completely in the ray optics picture, it can survive as a scarred mode with well-defined mode spacing in the spectrum and with emission directionality as seen in our experiment.

In summary, we investigated the lasing characteristics of asymmetrically deformed liquid jet and observed mode-order-dependent $Q$ spoiling of MDR modes. 
We could clearly identify persistent lasing modes with emission directionality for large deformation ($\eta >$ 12\%) beyond which the ray optics model does not allow any resonance modes. 
Our results can be explained by the scar, the wave natures of the unstable periodic orbits in chaotic ray dynamics.  On practical side, the present work provides a foundation for $Q$-tailoring and optimal deformation of ultrahigh-$Q$ microcavities for stronger and more directional light-emitting optoelectronic devices.

This work is supported by Creative Research Initiatives of the Korean Ministry of Science and Technology.

\bibliographystyle{prsty}

\begin{figure}
\caption{Experimental setup for deforming a liquid jet by lateral gas flow.}
\label{setup}
\end{figure}

\begin{figure}
\caption{Spectrum of MDR lasing for various $\eta$.}
\label{spectrum}
\end{figure}


\begin{figure}
\caption{(a) Correlation between the amount of red shift of individual MDR modes and the deformation parameter $\eta$. (b)Degradation of $Q_{\rm leak}$ as a function of $\eta$.}
\label{eta-dependence}
\end{figure}

\begin{figure}
\caption{Poincare surface of section for (a) $\eta=$ 10\% and (b) $\eta=$ 14\%.}
\label{poincare sos}
\end{figure}

\begin{figure}
\caption{Calculated wave function for (a) resonant ($2\pi a/\lambda$= 45.055) and (b) non-resonant ($2\pi a/\lambda$= 45.000) cases. A plane wave is incident from the left.}
\label{scar}
\end{figure}

\begin{figure}
\caption{Emission directionality seen in the far-field distribution.}
\label{direction}
\end{figure}

\end{document}